\documentclass[12pt]{article}
\usepackage{subfigure}
\usepackage{url}
\usepackage{color}
\usepackage{xcolor}
\usepackage{graphicx}
\usepackage{amsmath}%
\usepackage{lipsum}
\usepackage{mathtools}
\usepackage{amssymb}
\usepackage[a4paper, left=2.5cm, right=2.5cm, top=3cm, bottom=3cm]{geometry} 
\usepackage{authblk}
\newcounter{bla}

\graphicspath{{./figs/}}
\newcommand{\LIBNAME}{pyDOF}

\title{\LIBNAME: a Python library for the design of discrete forward and inverse filters}

\author[a,b]{Z. Nikolaou \footnote{Corresponding author: ZachariasMNic@gmail.com}}
\author[b]{P. Domingo}
\author[b]{L. Vervisch}
\author[a]{D. Drikakis}

\affil[a]{Institute for Advanced Modelling and Simulation, University of Nicosia, Nicosia CY-2417, Cyprus.}
\affil[b]{CORIA-CNRS, Normandie Université, INSA de Rouen Normandie, 685 Av. de l’Université, 76800 Saint-Étienne-du-Rouvray, France.}

\begin{document}

\maketitle

\hrule

\begin{abstract}
In this work, we present \LIBNAME, a Python-based software library which provides a domain-specific framework for the design of symmetric, physical-space, forward as well as inverse discrete filters. \LIBNAME\ is based on a constrained optimisation framework developed in our previous work \cite{nikolaou_caf_2023, 2024_jfm_nikolaou}. This framework allows the user to impose a wide range of constraints on the discrete filter transfer-function such as monotonicity, positivity, value-fixing, gradient-smoothing etc. amongst many others. \LIBNAME\ additionally includes an adaptive filter stencil selection option, and a van Cittert-based inverse-filter design with a user-controlled reconstruction order. The filter coefficients are computed automatically, and saved to a plain text file which can be readily parsed by any programming language. \LIBNAME\ can be used to design a wide range of low-pass, high-pass, multi band-pass/band-stop etc. discrete filters. In addition, due to its generality and abstraction, \LIBNAME\ can be used to design specific filters for user-defined target filter transfer functions. Although developed primarily for application to computational fluid dynamics simulations, \LIBNAME\ can be used to design discrete filters for a wide range of signal processing applications.\\
\textit{GitHub: \url{https://github.com/znikolaou/pyDOF.git}}
\end{abstract}

\newpage

\section{Introduction}

Filtering is a fundamental process in modern signal processing which has diverse and numerous industrial applications in audio, image, and video processing, telecommunications, time-series analysis etc. \cite{1995_orfanidesBook, 1999_oppenheimBook}. Discrete filters are typically used to de-noise signals, smooth or enhance signals, but also to extract useful physical information from complex and large-scale data sets \cite{2013_compPhysComm_merino, 2014_compPhysComm_merino, 2025_compPhysComm_zhang}. Discrete filters are also used extensively in numerical simulations to damp out in space (or in time) undesirable wavenumber components of the solution thereby stabilising the numerical solver \cite{2024_compPhysComm_xie} but also for numerous pre-processing or post-processing operations. 

In this work, our interest originates from the application of discrete filters in Computational Fluid Dynamics (CFD) simulations, and specifically Large Eddy Simulations (LES). In explicit LES, discrete filters form the building blocks of numerous major turbulence and combustion models. Classic models such as the dynamic Smagorinsky model \cite{smagorinsky_mwr_1963}, the scale-similarity model \cite{bardina_1983}, and models of the mixed kind \cite{1996_theorCompFluidDyn_vreman} employ filtering operations to calculate critical model parameters. These filtering operations in LES are typically performed on 3D signals, and on computational meshes with millions of points/cells. As such, the filters must be accurate but at the same time compact (small stencil size) in order to reduce the computational cost associated with expensive 3D convolution operations in physical space. Discrete filters in LES also form the basis of novel models based on deconvolution/reconstruction \cite{2024_jfm_nikolaou, stolz_pof_1999, stolz_pof_2001}. In such models, reconstruction algorithms such as van Cittert iterations \cite{van_cittert_1931}, use iterative filtering operations to obtain estimates of the unfiltered fields from their filtered counterparts on the computational mesh. In turn, these estimates form the basis for developing turbulence and combustion models. Such deconvolution-based models have been developed and tested successfully both for non-reacting \cite{stolz_pof_1999, stolz_pof_2001, schlatter_ijhff_2004, san_ijcf_2015, 2018_jCompApplMath_maulik, 2021_pof_zelong, 2025_compMathAppl_boguslawski} and reacting flows \cite{mathew_proccomb_2002, domingo_proccomb_2015, 2018_aeron_nikolaou, nikolaou_ftc_2018, nikolaou_prf_2018, nikolaou_prf_2019, wang_cnf_2017, wang_cnf_2019, Domingo2020, 2022_cnf_data}. An alternative to using expensive iterative reconstruction algorithms, is to employ a suitably designed direct-inverse discrete filter. This approach can substantially reduce the computational cost associated with expensive iterative algorithms as shown in our previous work \cite{nikolaou_caf_2023}. 

The (forward) filters used in LES are typically low-pass and symmetric. Gaussian, Helmholtz, and Implicit (Pad\'{e}) filters are some popular choices \cite{lele_jcp_1992}. The transfer function of such filters, unlike the ideal spectral cut-off low-pass filter, is relatively smoother and positive over the resolved wavenumber range $kh \in [0, \pi]$ where $h$ is the mesh spacing. This is a useful attribute which makes them invertible for all wavenumbers on the mesh. In practice, the filtering and reconstruction operations are performed in physical space for a given time-step, and for every point/cell of the computational mesh. In contrast to using Fourier transforms to perform the convolution operations, this approach is more straightforward to implement, and to parallelise in CFD solvers \cite{2016_compPhysComm_municchi} while still accounting for a wide range of boundary conditions. Therefore, our interest is to obtain optimised filter coefficients for applying the convolution operations in physical space. 

The standard approach for obtaining the filter coefficients in the fluid mechanics community, is to use truncated Taylor-series expansions of the filtering operation. This leads to analytical filter expressions with the accuracy of the filter depending on the truncation order \cite{sagaut_ijnf_1999}. In general, the approximate filters obtained using this approach, although computationally very efficient (for low truncation orders), have poorer performance especially at higher wavenumbers and close the grid cut-off wavenumber $k h=\pi$. In addition, such filters are generally not consistent i.e. constant signals may be damped or enhanced \cite{nikolaou_caf_2023} which is undesirable. Ideally, we would like to ensure that $\hat{G}_d(kh=0)=1$ where $\hat{G}_d$ is the discrete filter transfer function. Furthermore, their transfer function may become negative for some of the resolved wavenumbers on the mesh. This leads to unstable reconstruction, and makes them unsuitable for deconvolution-based modelling in LES \cite{nikolaou_caf_2023}. It would therefore be extremely useful, and essential, to be able to design forward and inverse filters through a general, flexible, and quantitative process allowing the user fine-grain control over the filter transfer function a priori. 

In the signal processing community, numerous methods were developed throughout the years for designing discrete filters. Notable classic methods include the window method, and the Parks-McClellan approach \cite{1972a_ieee_parks, 1972b_ieee_parks, 1975_ieee_rabiner}. Window methods use a weight function (window) to essentially regularise the ideal impulse response of the filter which in turn results in a smoother transfer function. Popular choices include Hamming and Kaiser windows \cite{1974_procieee_kaiser} with both approaches being computationally very efficient. Kaiser windows additionally allow the user to set the transition width which is a very useful property in discrete filter design. The Parks-McClellan approach \cite{1972a_ieee_parks} is a substantially different method. It is an iterative method based on the Remez exchange algorithm. Although somewhat more expensive than the window method, filters obtained using the Parks-McClellan method are optimal in the sense that the maximum error between the desired and actual frequency responses (transfer function) is minimized, and exhibit an equi-ripple behavior in the pass-band and stop-band ranges-in turn this ensures that the error is bounded in these regions. Even though this approach also allows the user to set the transition width just like the Kaiser-window method, it does not allow explicit and independent control of the passband ripple, and the stop-band attenuation. Optimisation-based techniques for filter design have also been developed \cite{1988_ieeetrans_medlin}. In least-squares optimisation, the filter is optimum in the sense that the error is minimised between the actual and ideal frequency responses however the classic un-constrained approach does not allow any control of the pass-band ripple or the transition width. To overcome some of the drawbacks of the aforementioned approaches, later methods focused on employing constrained optimisation approaches. The METEOR code developed in the early 1990s \cite{1992meteorCode} is one of the earliest examples. In more recent studies, a wide range of different evolutionary algorithms have been employed as well to design a wide range of both explicit and implicit filters. These include for instance particle-swarm optimisation \cite{2003_appOpt_zhou}, genetic algorithms \cite{1982_icassp_etter, 1991_ieee_suckley, 1986_ieee_ng}, and artificial bee-colony algorithms \cite{2009_jfi_karaboga}. Most classic filter design methods (window methods, Parks-McClellan), are nowadays readily available in commercial software such as Matlab's signal-processing utility but also in Python's free scipy signal-processing library. More complex methods based on evolutionary algorithms may not be readily available and/or require substantial experience and external software libraries to use them effectively. In addition, the computational cost of evolutionary algorithms may be significantly larger than that of classic methods. It is also important to note at this point that the majority of studies in the literature have focused only on the design of forward filters while our specific interest is to be able to design both forward and inverse filters using a single unified framework. In addition, we would like to be able to control the reconstruction order explicitly as this determines the range of wavenumbers being recovered on the computational mesh. In order to address these as well as other issues, a constrained optimisation framework for designing both forward and inverse discrete filters was developed in our earlier work \cite{nikolaou_caf_2023}. The inverse filter design is based on van Cittert regularisation with the reconstruction order being controlled explicitly via a single user-defined parameter $N$-the number of iterations in the traditional form of the approach. This framework was then used to develop forward and inverse discrete filters, and was extensively validated by filtering and reconstructing 1D but also full 3D turbulent fields on large-scale computational meshes with millions of points \cite{2024_jfm_nikolaou}. 

In this work, this framework has been extended, refined, and implemented in the \LIBNAME\ library. In contrast to most commercial or freely-available filter design tools, \LIBNAME\
can be used to design forward and inverse filters. In addition, \LIBNAME\ allows the user to select amongst a wide range of constraints and by doing so explicitly control the characteristics of the discrete filter transfer function. \LIBNAME\ additionally has the capability of computing the filter stencil size adaptively so as to achieve a target error metric. These, as well as other capabilities are thoroughly discussed in the text which follows. It is important to note at this point that \LIBNAME\ is a comprehensive and domain-specific framework for developing discrete forward and inverse filters. The user may well employ alternative optimisation algorithms whilst still enjoying the benefits of using \LIBNAME\ which is abstract and not bound to any specific optimiser. Section \ref{sec:mathBackground} gives a short description of the theoretical background, section \ref{sec:implementation} describes the high-level classes and methods, and section \ref{sec:validation} demonstrates the use of \LIBNAME\ for developing a wide range of forward and inverse filters.  

\section{Mathematical background} \label{sec:mathBackground}

\subsection{Filtering and reconstruction}

A Cartesian coordinate system will be used throughout the analysis. Since 3D filters in CFD are typically obtained using dimensional splitting i.e. applying 1D filters in each coordinate direction \cite{sagaut_ijnf_1999}, we restrict our analysis to 1D as per our previous studies \cite{nikolaou_caf_2023, 2024_jfm_nikolaou}. Consider a regular 1D mesh with grid-points given by $x_i$=$ih$, where $h$ is the mesh spacing, $i \in [0,N_x-1]$, and $N_x h=L$. In the majority of CFD applications we are interested in symmetric and explicit filters. Therefore, for a signal $u_i=u(x_i)$ on the mesh, the forward discrete filtering operation is defined by, 

\begin{equation}\label{eq:filt_discr}
 G_d \circledast u_i \coloneqq \bar{u}_i=\sum_{l=-M_F}^{M_F} g_l u _{i+l}
\end{equation}

\noindent where $G_d$ is the discrete filter kernel, $g_l$ are the forward filter coefficients, and $N_s=(2M_F+1)$ is the stencil size of the filter. By analogy, we can define an inverse (reconstruction) filtering operation as follows, 

\begin{equation*}
V_d \circledast \bar{u}_i \coloneqq {u^*_i}=\sum_{l=-M_{I}}^{M_{I}}b_l\bar{u}_{i+l}
\end{equation*}

\noindent where $V_d$ is the inverse filter kernel, $u^*_i$ is the reconstructed signal, $b_l$ are the inverse filter coefficients, and $N_s=(2M_{I}+1)$ is the stencil size of the inverse filter. Note that the inverse filter stencil size may not necessarily be equal to the stencil size of the forward filter. Also note that the inverse filtering operation is performed on the filtered field, and for a perfect reconstruction, $u^*_i=u_i \hspace{0.25cm} \forall i \in [0, N_x-1]$.

The frequency response (transfer function) of the filtering operation can be obtained by performing a von Neumann (Fourier) analysis. Consider a discrete periodic signal,

\begin{equation}\label{eq:periodicSignal}
\phi_i=\sum_{r=0}^{N_x-1}a_re^{jk_rx_i}
\end{equation}

\noindent where $k_r=2\pi r/L$ is the corresponding wavenumber. Filtering this sinusoid using Eq. \ref{eq:filt_discr} we obtain, 

\begin{equation*}
\bar{\phi}_i=\sum_{r=0}^{N_x-1}a_re^{jk_rx_i}\hat{G}_d(k_rh)
\end{equation*}

\noindent where 

\begin{equation*}
\hat{G}_d(k_rh)=\sum_{l=-M_F}^{M_F}g_le^{j k_rh l}=g_0+2\sum_{l=1}^{M_F}g_l cos \left( k_rh  l \right)
\end{equation*}

\noindent is the transfer function (frequency response) of the discrete filter. Note that due to the symmetry of the filtering operation, the transfer function is always real. Provided a target filter transfer function in the continuous domain $\hat{T}$ is known, the objective is to find coefficients $g_l$ such that $\hat{G}_d$ best matches $\hat{T}$ in some sense which we will define later on. In the case of inverse filters, a target transfer function is also required. The most straightforward way to obtain a regularised inverse target transfer function is by using van Cittert reconstruction as explained in the section which follows.  

\subsection{van Cittert reconstruction} \label{sec:convergence_vancit}

The van Cittert algorithm \cite{van_cittert_1931} is a fundamental linear iterative reconstruction algorithm which recovers an estimate of the unfiltered field, $u^*_i$, using  successive filtering operations on the filtered field $\bar{u}_i$. In the discrete case, the algorithm is applied point-wise to every point in the computational domain as follows, 

\begin{equation}\label{eq:van_cittert_discr}
u^{*N+1}_i=u^{*N}_i+b(\bar{u}_i-G_d \circledast u^{*N}_i)
\end{equation}

\noindent where $u^{*0}_i=\bar{u}_i$, $N$ is the number of iterations defining the reconstruction order, and the parameter $b$ is typically taken to equal unity. For a detailed description of the convergence properties of Eq. \ref{eq:van_cittert_discr}, we refer the reader to \cite{nikolaou_caf_2023}. In order to determine the response of the reconstruction process to periodic signals, we follow the same approach as in the previous section. Specifically, inserting Eq. \ref{eq:periodicSignal} in Eq. \ref{eq:van_cittert_discr} and expanding, we find after some algebraic manipulation that, 

\begin{equation*}
{\phi ^* _i} ^{N}=\underbrace{\sum_{r=0}^{N_x-1}a_r e^{jk_r ih}}_{\phi _i}+\underbrace{\sum_{r=0}^{N_x-1} \left( 1-b\hat{G}_d(k_rh) \right)^N \left( \hat{G}_d(k_rh)-1 \right)a_r e^{jk_rih}}_{error}
\end{equation*}
  
\noindent The first term is the original field, and the second term is the reconstruction error. The above  can also be written as, 

\begin{equation*}
{\phi ^* _i} ^{N}={\sum_{r=0}^{N_x-1}a_r \left(1-\left( 1-b\hat{G}_d(k_rh) \right)^{N} \left( 1-\hat{G}_d(k_rh \right) \right) e^{jk_r ih}}=\sum_{r=0}^{N_x-1}a_r\hat{Q}_d^N(k_rh)e^{jk_r ih}
\end{equation*}
  
\noindent where $\hat{Q}_d^N(k_rh)$ is the transfer function of the reconstructed signal. In the case $b=1$ (which is a convenient and popular choice) we have, 

\begin{equation}\label{eq:regInverseTarget}
{\hat{Q}^N}_d(k_rh)=1-\left( 1-\hat{G}_d(k_rh) \right)^{N+1} 
\end{equation}
 
\noindent Note that if $\hat{G}_d(k_rh)=0$ for some wavenumber, then $\hat{Q}_d(k_rh)=0$ always i.e. this wavenumber component can never be recovered. It is straightforward to show that provided $\hat{G}_d \in (0,1]$ then $\hat{Q}_d \rightarrow 1$ as $N \rightarrow \infty$ i.e. all wavenumbers on the mesh are recovered thereby restoring the original signal. For lower values of $N$, the filtered wavenumbers are partially recovered-\LIBNAME\ allows the user to set this. For a given reconstruction order $N$, ${\hat{Q}^N}_d$ can be taken to be the target for $\hat{V}_d \hat{G}_d$ since this is the transfer function of the discrete reconstructed signal which is obtained by multiplying the inverse transfer function $\hat{V}_d$ with the forward transfer function $\hat{G}_d$ \cite{nikolaou_caf_2023}. Another option, is to use directly the regularised inverse as the target. This is is simply given by $\hat{Q}^N_d/\hat{G}_d$.  

\subsection{Formulation of the optimisation problem}\label{sec:formOpt}

\begin{table}[h!]
\centering
\small
\begin{tabular}{cccc}
Filter class   & $\hat{T}(k)$ & $k_ch$ & User-defined params \\
\hline
Gaussian  & $e^{\frac{-{(\Delta/h)} ^2(kh)^2}{24}}$  & $\frac{2}{\Delta /h}\sqrt{6ln(2)}$ & $\Delta /h $\\
Helmholtz  & $\frac{1}{1+\lambda ^2 k^2}$ & $\frac{1}{\lambda /h}$ & $\lambda / h$  \\
Implicit & $\frac{ (1/2+a)(1+cos(kh))}{1+2acos(kh)}$ & $arccos(-2a)$ & $a$ \\
Butterworth    & $\frac{1}{\sqrt{1+3\left( \frac{k}{k_c} \right)^{2n}}}$  & Variable & $n, k_c$ \\
Sharp & Low-pass, high-pass, & Variable & kbounds []\\ & multi-pass, multi-stop. & & \\
\hline
\end{tabular}
\caption{The different filter classes available in \LIBNAME. The \textit{Sharp} class can be used to define a wide range of ideal filters by providing the list of band edges. All these classes inherit the \textit{Filter} class, and the user may also implement their own custom-made target transfer functions by following the inheritance pattern.}
\label{tbl:filters}
\end{table}

\LIBNAME\ allows the user to define two different error metrics for the optimiser: mean-squared error, and mean absolute error. For a general discrete transfer function $\hat{G}_d(k_rh; g_l)$ where $g_l$ are the discrete filter coefficients, and for a target transfer function $\hat{T}(k_rh; \underline{\chi})$, where $\underline{\chi}$ are any parameters affecting the shape of $\hat{T}$, the two error metrics are defined as follows, 

\begin{equation*}
E_{sq} \left( \hat{G}_d, \hat{T} \right)=\frac{1}{N_k}\sum_{k_rh=0}^{\pi}{\left( \hat{G}_d(k_rh;g_l)-\hat{T}(k_rh; \underline{\chi}) \right)}^2
\end{equation*}

\begin{equation*}
E_{abs} \left( \hat{G}_d, \hat{T} \right)=\frac{1}{N_k}\sum_{k_rh=0}^{\pi}{\left| \hat{G}_d(k_rh;g_l)-\hat{T}(k_rh; \underline{\chi}) \right|}
\end{equation*}

\noindent where $N_k$ is the number of points used to sample the wavenumber space. This parameter is preset in the library but the user may well change it. \LIBNAME\ includes a number of different target filter transfer functions $\hat{T}$ as shown in Table \ref{tbl:filters}. In addition to the above two error metrics, we define the following gradient smoothing terms for each of the above two error metrics, 

\begin{equation*}
S_{sq}(\hat{G}_d)=\frac{1}{N_k}\sum_{k_rh=0}^{\pi} \left| \nabla \hat{G}_d(k_rh;g_l) \right|^2
\end{equation*}

\begin{equation*}
S_{abs}(\hat{G}_d)=\frac{1}{N_k}\sum_{k_rh=0}^{\pi} \left| \nabla \hat{G}_d(k_rh;g_l) \right|
\end{equation*}

\noindent For forward filters having a discrete transfer function $\hat{G}_d$, the optimisation problems for the two different error definitions are formulated as follows,   

\begin{equation*}
 \underset{g_l}{\text{arg min}} \left( E_{sq} \left( \hat{G}_d(k_rh; g_l), \hat{T}(k_rh; \underline{\chi} ) \right)+ \lambda S_{sq}(\hat{G}_d) \right)\\
\end{equation*}

\begin{equation*}
 \underset{g_l}{\text{arg min}} \left( E_{abs} \left ( \hat{G}_d(k_rh; g_l), \hat{T}(k_rh; \underline{\chi} ) \right )+ \lambda S_{abs}(\hat{G}_d) \right)\\
\end{equation*}

\noindent where $\lambda$ is a positive smoothing parameter. By default $\lambda=0$ however the user may set $\lambda$ to any positive value to smooth the discrete filter transfer function accordingly. For inverse filters, the formulations are somewhat different. Having solved the optimisation problem for a discrete forward filter, and having obtained the forward filter coefficients $g_l$, the objective is to obtain an inverse filter with a transfer function $\hat{V}_d(k_rh;b_l)$ such that applying the inverse filter to a signal filtered using a forward filter with transfer function $\hat{G}_d(k_rh;g_l)$ we obtain an estimate of the original signal. Therefore, the optimisation problems for each of the two different error metrics are formulated as follows,

\begin{equation*}
 \underset{b_l}{\text{arg min}} \left( E_{sq} \left( \hat{V}_d(k_rh; b_l)\hat{G}_d(k_rh; g_l), \hat{Q}^N_d(k_rh;g_l)\right)+ \lambda S_{sq}(\hat{V}_d) \right)\\
\end{equation*}

\begin{equation*}
 \underset{b_l}{\text{arg min}} \left( E_{abs} \left( \hat{V}_d(k_rh; b_l)\hat{G}_d(k_rh; g_l), \hat{Q}^N_d(k_rh;g_l)\right)+ \lambda S_{abs}(\hat{V}_d) \right)\\
\end{equation*}

\noindent Note that the target for the inverse filter design is $\hat{Q}^N_d$ which is the regularised discrete inverse based on van Cittert iterations as given in Eq. \ref{eq:regInverseTarget}. \LIBNAME\, also allows the user to use $\hat{Q}^N$ i.e. the regularised inverse obtained using the continuous target $\hat{G}$ instead of $\hat{G}_d$. 

\begin{table}[h!]
\centering
\footnotesize
\begin{tabular}{ccc}
Constraint & \LIBNAME\ keyword  & Default value\\
\hline
$\hat{G}_d(k_rh)=1$   & fixOne  & None\\
$\hat{G}_d(k_rh)=0$   & fixZero & None\\
$u^i_l\leq G_d(k_rh) \leq u^i_u$ for $ k \in [k^i_l, k^i_u]$ & luBounds & None\\
$0 \leq \hat{G}_d(k_rh) \leq 1$ $\forall k_rh \in (0, \pi)$ & zeroToOne & False\\
$0 \leq \hat{G}_d(k_rh) $ $\forall k_rh \in (0, \pi)$  & positive & False\\
$d\hat{G}_d / {dk} \geq 0$ $\forall k_rh \in [0, \pi]$  & monotonePos & False\\ 
$d\hat{G}_d /{dk} \leq 0$ $\forall k_rh \in [0, \pi]$ & monotoneNeg & False \\
  & adaptive & True \\
\hline                                                                                                                                                  
\end{tabular}
\caption{The different optimisation constraints and control keywords available in \LIBNAME.}
\label{tbl:constraints}
\end{table}

The optimisation problems above can be solved in a fully unconstrained approach or subject to a variety of linear constraints as shown in Table \ref{tbl:constraints}. The first three constraints in Table \ref{tbl:constraints} can be  used to fix/bound the value of the discrete filter transfer function. The rest can be used to restrict the transfer function to be positive, between 0 and 1, and also constrain the gradient of the transfer function to be monotonic positive or negative. The adaptive constraint keyword instructs the optimiser to compute the stencil size dynamically. Additional keywords allow the user to set the optimisation error, the regularisation parameter $\lambda$ etc. as per the examples in the \textit{./examples/} directory.

\section{Implementation}\label{sec:implementation}

In this section, a brief overview of \LIBNAME's main classes and methods is given. For more detailed descriptions of how to set-up an optimisation problem and compute filter coefficients, we refer the reader to the ./examples/ directory. \\

\noindent \textit{Wavenumber:} \\

\noindent This is the class for defining the wavenumber object. By default, the range is set to $kh \in [0, \pi]$, and the user can provide the number of points for the wavenumber stencil. \\

\noindent \textit{Filter:}\\

\noindent This is the base class for all filters. All filters must inherit this classm and follow the inheritance pattern. To create an object of a filter, the user must provide a \textit{Wavenumber} object as well as the filter parameter(s). For example, to create a \textit{Gaussian} filter with a parameter $\Delta/h=4$ on a 1000-point mesh for the wavenumbers, \\

\begin{center}
\textit{waveNoOb=Wavenumber(nk=1000)}\\
\textit{gaussianFilterOb=Gaussian(waveNoOb, 4)}\\
\end{center}

\noindent To obtain the transfer function we simply call, \\

\begin{center}
\textit{trf=gaussianFilterOb.getTransferFunction()} \\
\end{center}

\noindent \textit{DiscreteFilter:}\\

\noindent This is the base class for defining a discrete filter. To create an object of this class, the user must supply a \textit{Wavenumber} object, the discrete filter stencil size $N_s$, and define whether the filter is a forward one (\textit{isForward=True}) or an inverse one (\textit{isForward=False}). For example,  to create a discrete forward filter on a 21-point stencil, \\

\begin{center}
\textit{dfltOb=DiscreteFilter(waveNoOb, nStencil=21, isForward=True)}\\
\end{center}

\noindent This will create the object only. To actually get the discrete filter transfer function, the user must first provide the half-stencil array of filter coefficients i.e. $g_0$, $g_1$, ..., $g_{10}$ as follows, \\

\begin{center}
\textit{dfltOb.setCoeffs([$g_0$, $g_1$, ..., $g_{10}$])}\\
\end{center}

\noindent To obtain the transfer function of the discrete filter we do, \\

\begin{center}
\textit{trfd=discreteFilterOb.getTransferFunction()}\\
\end{center}

\noindent \textit{ReconstructedTransferFunction:}\\

\noindent This class can be used to define an object for the reconstructed transfer function. This can be either $\hat{Q}^N$ based on the filter in physical space, or $\hat{Q}^N
_d$ i.e. based on the discrete filter. The reconstructed transfer function $\hat{Q}^N$ or $\hat{Q}^N_d$ will be the target $\hat{T}$ for obtaining the inverse discrete filter. For example, suppose we have calculated the coefficients of the forward discrete filter, and defined our discrete filter object as above. To obtain $\hat{Q}_d^{100}$ we first define the object as follows, \\

\begin{center}
\textit{recFuncOb=ReconstructedTransferFunction(dfltOb, nVanCittert=100)}\\
\end{center}

\noindent and to get the values of the reconstructed transfer function, \\

\begin{center}
\textit{qtrf=recFuncOb.getTransferFunction()}\\
\end{center}

\noindent To define an object for a reconstructed transfer function based on the forward transfer function in physical space, we additionally supply a \textit{Filter} object as follows,\\

\begin{center}
\textit{recFuncOb=ReconstructedTransferFunction(dfltOb, fltOb, nVanCittert=100)}\\
\end{center}

\noindent Then, when we invoke \textit{recFuncOb.getTransferFunction()}, we will obtain $\hat{Q}^{100}$ instead of $\hat{Q}_d^{100}$, and this will be the target for the optimiser instead. \\

\noindent \textit{Optimiser:}\\

\noindent This is the base class for setting up the optimisation problem. The user must provide a target transfer function $\hat{T}$ which is an instance of either \textit{Filter} or \textit{ReconstructedTransferFunction}, a \textit{DiscreteFilter} object as well as a range of other optimisation parameters and constraints. The available constraints are shown in Table \ref{tbl:constraints}. For detailed examples, we refer the reader to the \textit{./examples/} directory. An important parameter to set is the \textit{adaptive=True} or \textit{adaptive=False} switch. By default, this is set to \textit{True} in which case the optimiser will adjust the stencil size of the discrete filter ($M$) until the adaptive optimiser error is reached. If set to \textit{False}, the optimisation problem will be solved on a fixed value of $M$ as defined in the \textit{DiscreteFilter} object. For instance, to set up an optimisation problem for computing forward filter coefficients for a Gaussian filter as defined above using a fixed stencil size we do,

\begin{center}
\textit{optOb=Optimiser(gaussianFilterOb, discreteFilterOb, adaptive=False)}
\end{center} 

\noindent Then we run the optimiser using, 

\begin{center}
\textit{optOb.run()}
\end{center}

\noindent The optimiser will solve the optimisation problem and update the discrete filter coefficients in the provided \textit{DiscreteFilter} object. To obtain the half-filter stencil size coefficients we invoke,  

\begin{center}
\textit{optOb.get()}
\end{center}

\noindent or to obtain the full set of coefficients

\begin{center}
\textit{discreteFilterOb.getAllCoeffs()}
\end{center}

\section{Validation}\label{sec:validation}

\subsection{Low-pass forward filters}

\begin{figure}[h!]
\subfigure[]{
\includegraphics[width=0.55\textwidth, trim=3.0cm 0.0cm 0.0cm 0.0cm]{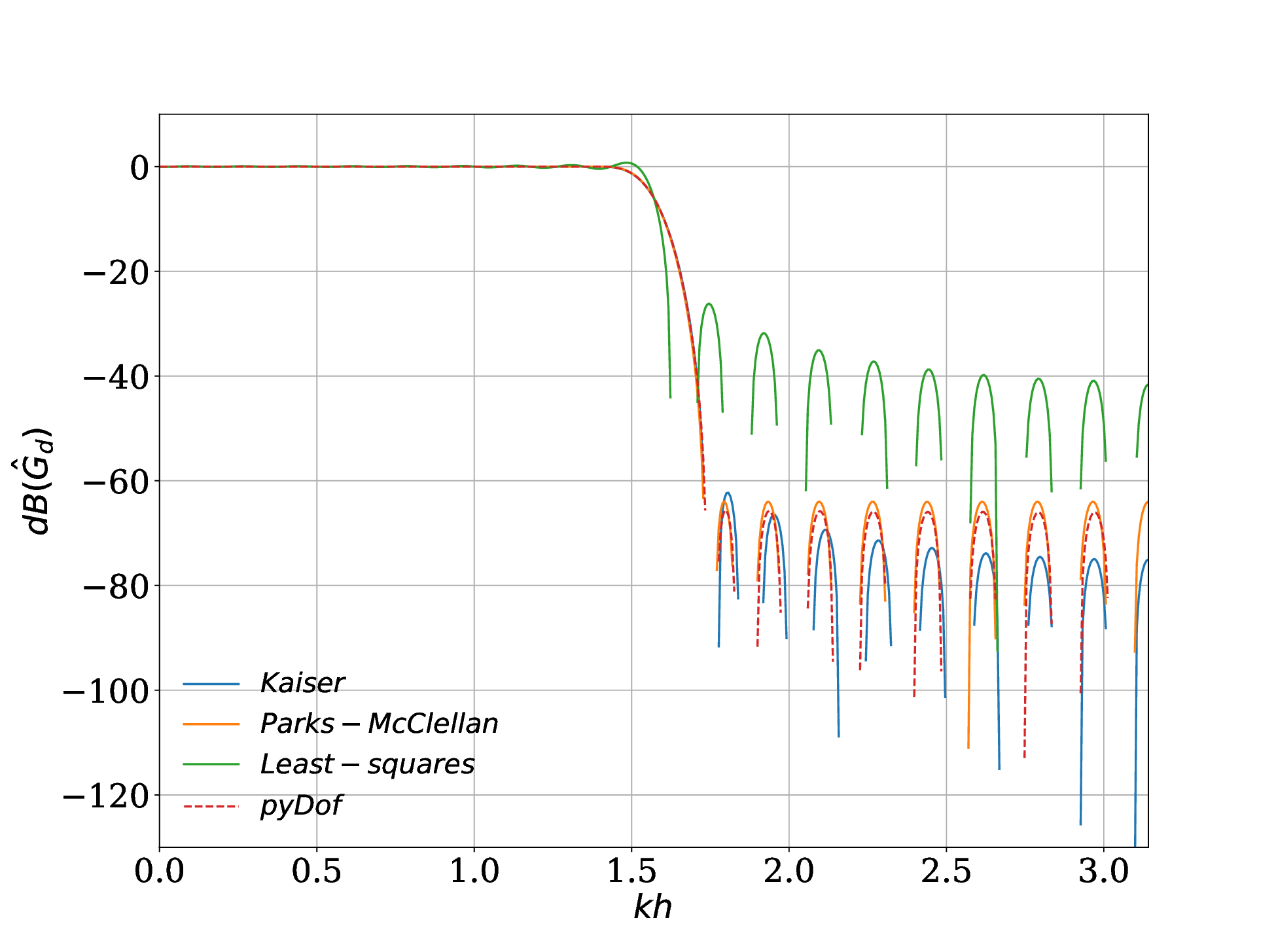}
}
\subfigure[]{
\includegraphics[width=0.55\textwidth, trim=3.0cm 0.0cm 0.0cm 0.0cm]{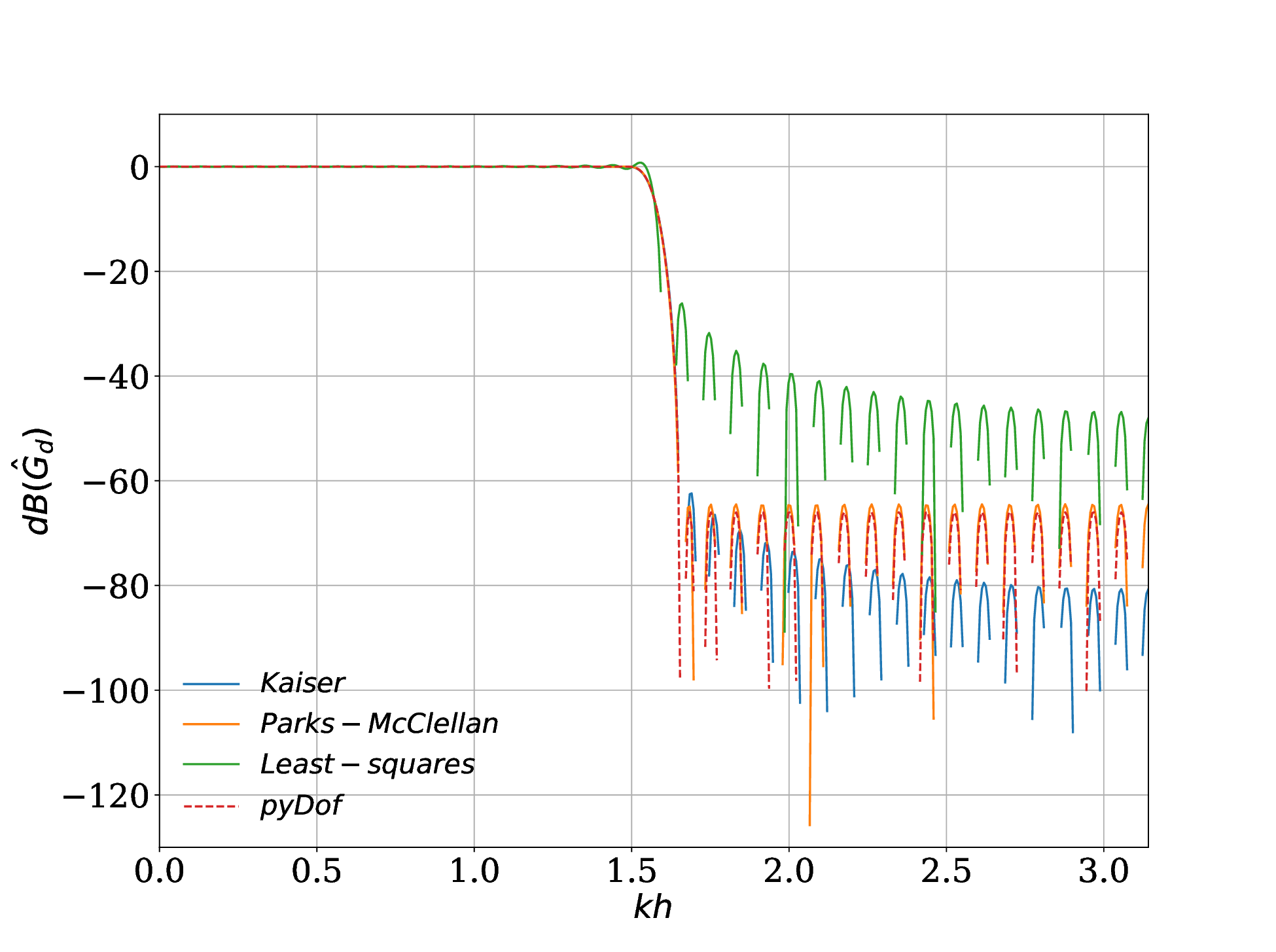}
}
\subfigure[]{
\includegraphics[width=0.55\textwidth, trim=3.0cm 0.0cm 0.0cm 0.0cm]{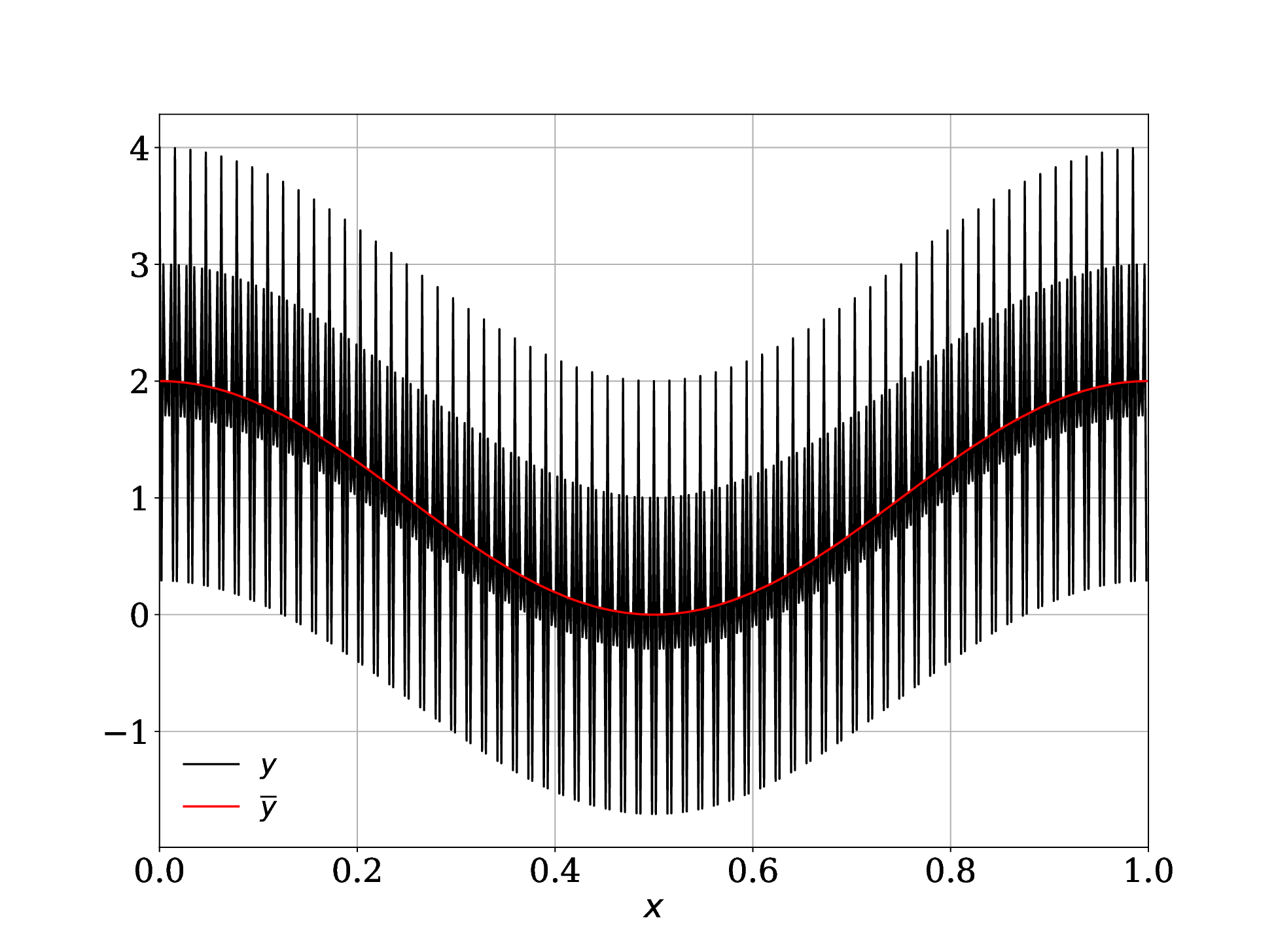}
}
\caption{Discrete filter transfer functions obtained using different methods for a low-pass filter: (a) $N_s=73$, (b) $N_s=145$. (c) shows the original signal $y$ (Eq. \ref{eq:lowPassTestSignal}), and the filtered signal $\bar{y}$ using the low-pass filter designed using \LIBNAME-its transfer function is shown in (b). }
\label{fig:methodsComparison}
\end{figure}

As a first step to validate \LIBNAME, we calculate discrete filter coefficients for an ideal low-pass filter. The stop-band attenuation $A$ is fixed at 60 dB, and we consider two different transition widths $\delta k$ of 0.1$\pi$ and $0.05\pi$. Using the standard filter design formula $N_s=1+D/(k_s/ \delta k)$ \cite{1995_orfanidesBook}, where $D=(A-7.95)/14.36$, we find the corresponding filter stencil sizes to be 73, and 145 points respectively for each transition width. The ideal low-pass filter cut-off wavenumber is set to $k_ch=\pi/2$. The performance of \LIBNAME\ is benchmarked against three different well-known methods in the literature: (i) using a Kaiser window, (ii) using the Parks-McClellan approach, and (iii) using least-squares optimisation. For the constrained optimisation approach implemented in \LIBNAME, in order to control the ripple in the pass-band and stop-band ranges, we set an luBounds constraint of $d=$0.0005 over the pass-band and stop-band ranges. Specifically, we set $1-d \leq \hat{G}_d(k_rh) \leq 1+d$ for $k_rh \in[0, kc_h-\delta k /2]$, and $-d \leq \hat{G}_d(k_rh) \leq d$ for $k_rh \in[kc_h+\delta k /2, \pi]$. In addition, to ensure consistency, we set $\hat{G}_d(0)=1$, and $\hat{G}_d(\pi)=0$. It is important to note that such constraints cannot be implemented using the standard methods above. The consequence of this is that the standard methods will not attenuate completely wavenumbers at the grid cut-off, and will not leave constant signals completely unaffected. The relevant Python script with the implemented optimisation problem (and all other classical methods) for this validation case is called \textit{mainLowPass.py} in the \textit{./examples/} directory.

\begin{table}[h!]
\centering
\footnotesize
\begin{tabular}{lll}
Low-pass, $N_s$=73         & MSQE       & Max \% overshoot in passband \\
\hline
Window-Kaiser                  & 4.9091E-03 & 1.2040E-01                   \\
Parks-McClellan                & 4.7680E-03 & 6.3178E-02                   \\
Least-squares                  & 2.7969E-03 & 8.9986E+00                   \\
pyDof                          & 4.8019E-03 & 5.0000E-02                   \\
Low-pass, $N_s$=145        &            &                              \\
\hline
Window-Kaiser                  & 2.4460E-03 & 1.1390E-01                   \\
Parks-McClellan                & 2.3643E-03 & 5.9515E-02                   \\
Least-squares                  & 1.3816E-03 & 9.0177E+00                   \\
pyDof                          & 2.3999E-03 & 5.0000E-02                   \\
Multi pass-band, $N_s$=73  &            &                              \\
\hline
Window-Kaiser                  & 1.9682E-02 & -2.1934E-04                  \\
Parks-McClellan                & 1.9425E-02 & 3.1631E-02                   \\
Least-squares                  & 1.1181E-02 & 7.5010E+00                   \\
pyDof                          & 1.9349E-02 & 3.0000E-02                   \\
Multi pass-band, $N_s$=145 &            &                              \\
\hline
Window-Kaiser                  & 9.8094E-03 & 1.8251E-01                   \\
Parks-McClellan                & 9.4267E-03 & 5.7983E-02                   \\
Least-squares                  & 5.2250E-03 & 8.5055E+00                   \\
pyDof                          & 9.6516E-03 & 5.0000E-02                  \\
\hline
\end{tabular}
\caption{Error metrics for the different filter design methods, and the different low-pass and multi-pass filters.}
\label{tbl:trfErrors}
\end{table}

Figures \ref{fig:methodsComparison} (a) and (b) show the transfer functions of the discrete filters obtained using the different methods while table \ref{tbl:trfErrors} shows the corresponding errors and performance metrics with respect to the ideal filter. As expected, least-squares optimisation leads to a transfer function which has  the lowest mean-squared error, and which is ``sharpest" around the transition point. This method however gives rise to a large overshoot around the cut-off wavenumber due to the well-known Gibbs phenomenon. In addition, the pass-band and stop-band ripples are substantial throughout the pass-band and stop-band ranges. The well-established Kaiser window and Parks-McClellan methods perform reasonably well as expected. These methods essentially sacrifice some of the resolution (larger transition width) in comparison to least-squares optimisation to produce lower ripples in the pass-band and stop-band. Of the two methods, Parks-McClellan has the lower (and uniform) overshoot in the pass-band but slightly larger in the stop-band. The constraint-based optimisation approach using \LIBNAME\ also performs remarkably well in comparison to the classic methods. It is important to note that despite the additional constraints imposed, the obtained transfer function results in similarly good error metrics both for the pass-band and stop-band ranges as shown in Table \ref{tbl:trfErrors}. In fact, the 0.05\% overshoot in the pass-band is a result of the 0.0005 ripple constraint which we imposed and which also remains uniform in the pass-band and stop-band. For the larger stencil size, \LIBNAME\ also performs well resulting in a sharper transfer function, and much closer to the ideal one as expected. This is also reflected in the lower errors as shown in Table \ref{tbl:trfErrors}. It is important to note that the overshoot for the Kaiser-Window method spikes near the cut-off wavenumber, and gradually reduces away from it. For this reason, even though the error in the stop-band range appears to be lower for this method, the overall mean-squared error remains of a similar magnitude as for the other two methods (Parks-McClellan, \LIBNAME). In terms, of computational time, the Kaiser window method is the cheapest, followed by Parks-McClellan, and the optimisation approaches.  

In order to further test the low-pass filter obtained using \LIBNAME\ in physical space as well, we consider a periodic signal $y_i$ in the domain $x_i=[0,1]$ where $i \in [0, N_x-1]$, and $N_x=512$. The signal is composed of two wavenumber components in the pass-band, and two in the stop-band as follows, 

\begin{equation}\label{eq:lowPassTestSignal}
y=1+cos \left(\frac{2\pi}{N} i \right)+cos \left(\frac{2\pi (3N/8)}{N} i \right)+cos\left(\frac{2\pi (N/2)}{N} i \right)
\end{equation}

\noindent Note that the grid cut-off wavenumber is also included. The coefficients obtained using \LIBNAME\ for the low-pass filter and for the larger stencil size ($N_s=145$) are used to filter $y$. The results are shown in Fig. \ref{fig:methodsComparison} (c). As expected, filtering removes the last two components in Eq. \ref{eq:lowPassTestSignal} since these fall in the stop-band. As a result, only the constant part, and the first sinusoid remain as clearly shown in Fig. \ref{fig:methodsComparison} (c). 

\subsection{Multi-pass forward filters}

\begin{figure}[h!]
\subfigure[]{
\includegraphics[width=0.55\textwidth, trim=3.0cm 0.0cm 0.0cm 0.0cm]{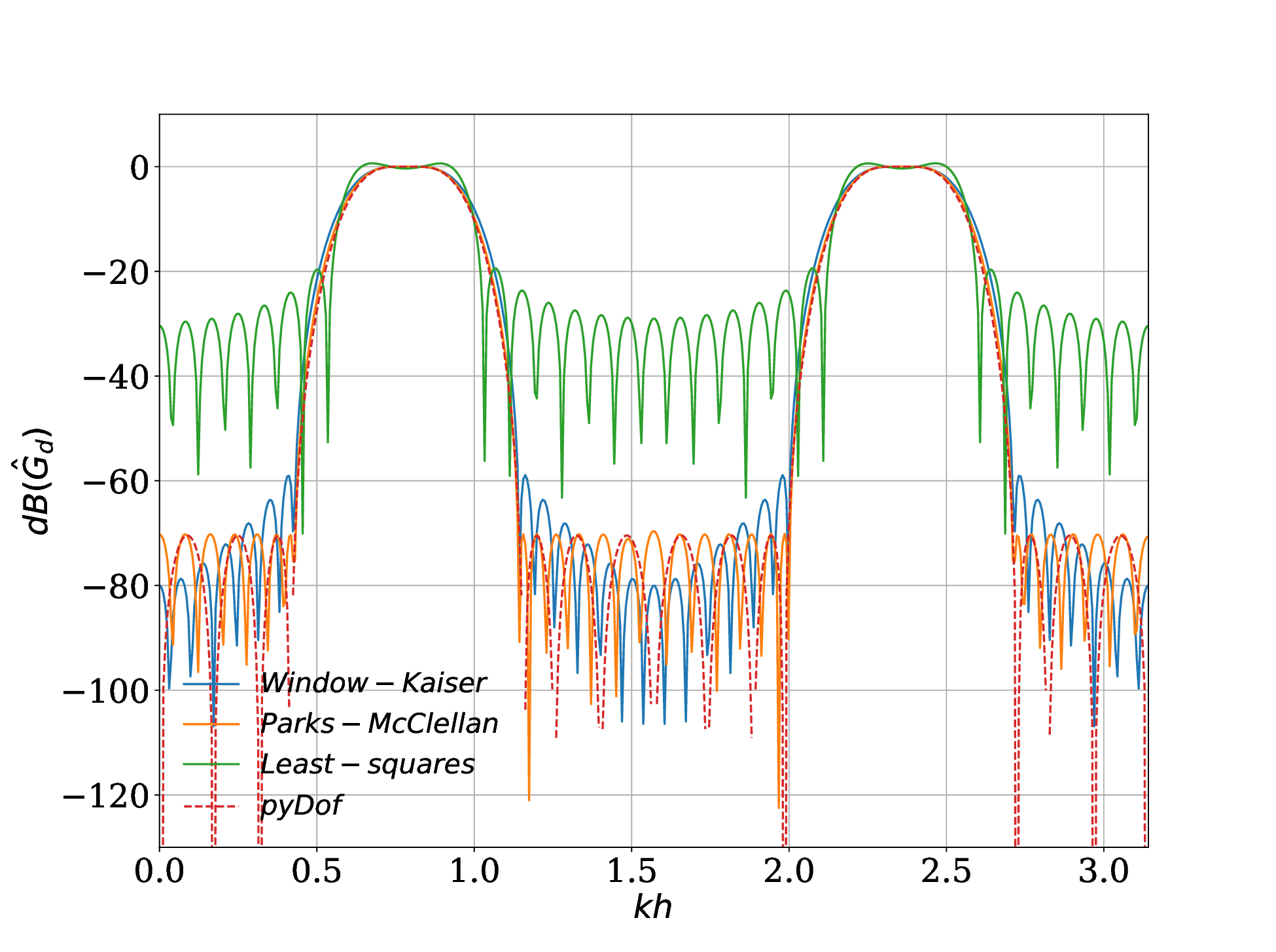}
}
\subfigure[]{
\includegraphics[width=0.55\textwidth, trim=3.0cm 0.0cm 0.0cm 0.0cm]{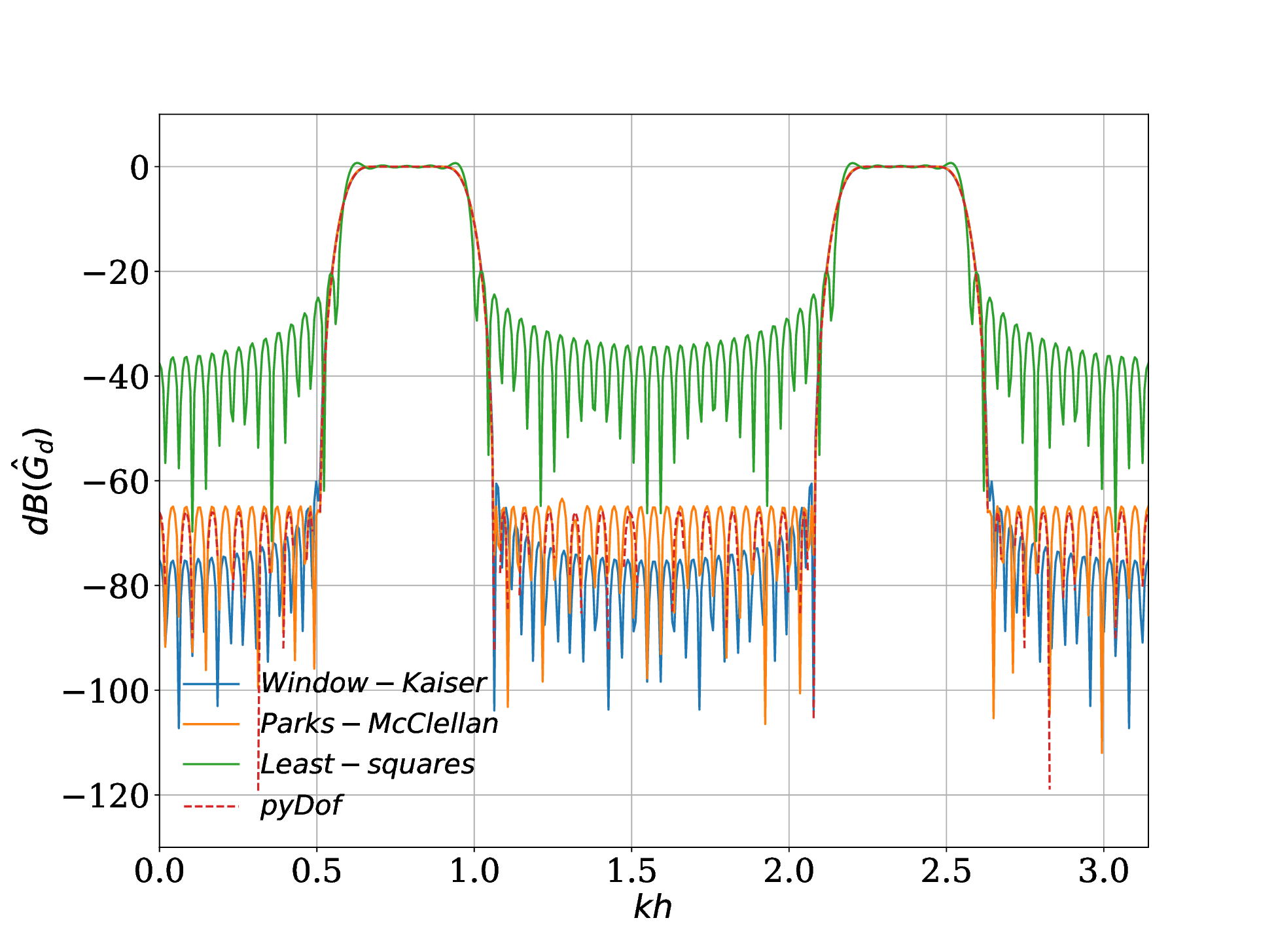}
}
\caption{Discrete filter transfer functions obtained using different methods for a multi pass-band filter: (a) $ns=73$, (b) $ns=145$.}
\label{fig:multiPassComparison}
\end{figure}

In a further validation step, \LIBNAME\ is used to design a  multi-pass forward filter. The transition bands are set in the ranges $k_ch=[\pi /4-\pi /16, \pi /4+\pi /16]$ and $k_ch=[3\pi /4-\pi /16, 3\pi /4+\pi /16]$. The ripple in the pass-band and stop-band ranges is fixed to 0.0003 for $ns=73$ and to 0.0005 for $ns=145$. In addition, to ensure that constant signals and wavenumber components at the grid cut-off are completely damped, we set $\hat{G}_d(0)=\hat{G_d}(\pi)=0$ when using \LIBNAME. Figure \ref{fig:multiPassComparison} shows the transfer functions of the discrete filters obtained using the different methods while Table \ref{tbl:trfErrors} shows the corresponding error metrics for each case. As with the low-pass filter, least-squares optimisation leads to transfer functions having the lowest errors, but with large ripples in the pass-band and stop-band ranges. The window and Parks-McClellan methods perform reasonably well with larger mean squared errors but with far smaller ripples in the pass-band and stop-band ranges. Using \LIBNAME\ we obtain similarly good results for both stencil sizes with the additional benefit of imposing additional constraints for constant signals and for wavenumber components at the grid cut-off. The relevant script for this validation case is called \textit{mainMultiPassBand.py} in the \textit{./examples/} directory.

\subsection{Gaussian forward and inverse filters}

\begin{figure}[h!]
\subfigure[]{
\includegraphics[width=0.55\textwidth, trim=3.0cm 0.0cm 0.0cm 0.0cm]{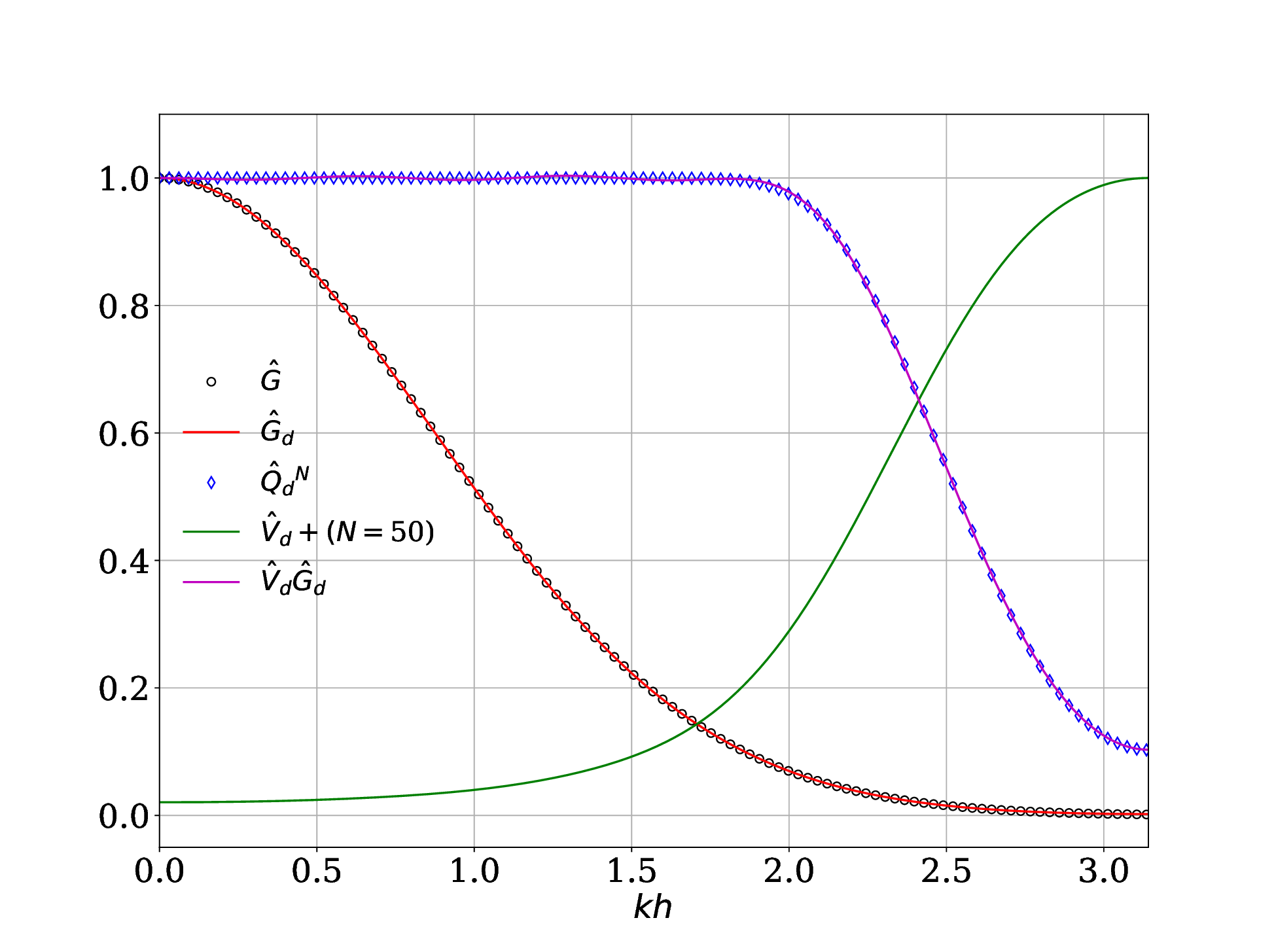}
}
\subfigure[]{
\includegraphics[width=0.55\textwidth, trim=3.0cm 0.0cm 0.0cm 0.0cm]{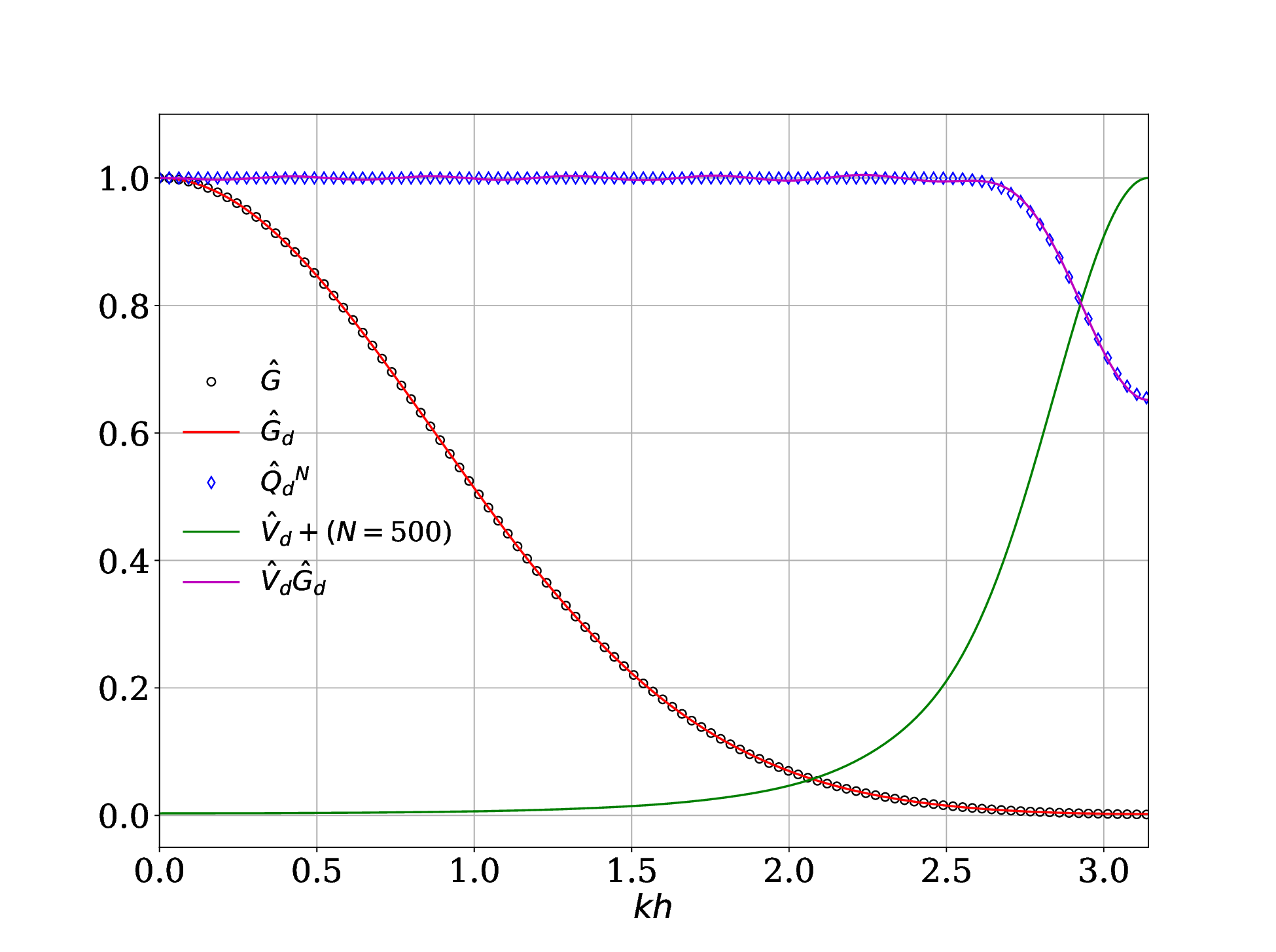}
}
\subfigure[]{
\includegraphics[width=0.55\textwidth, trim=3.0cm 0.0cm 0.0cm 0.0cm]{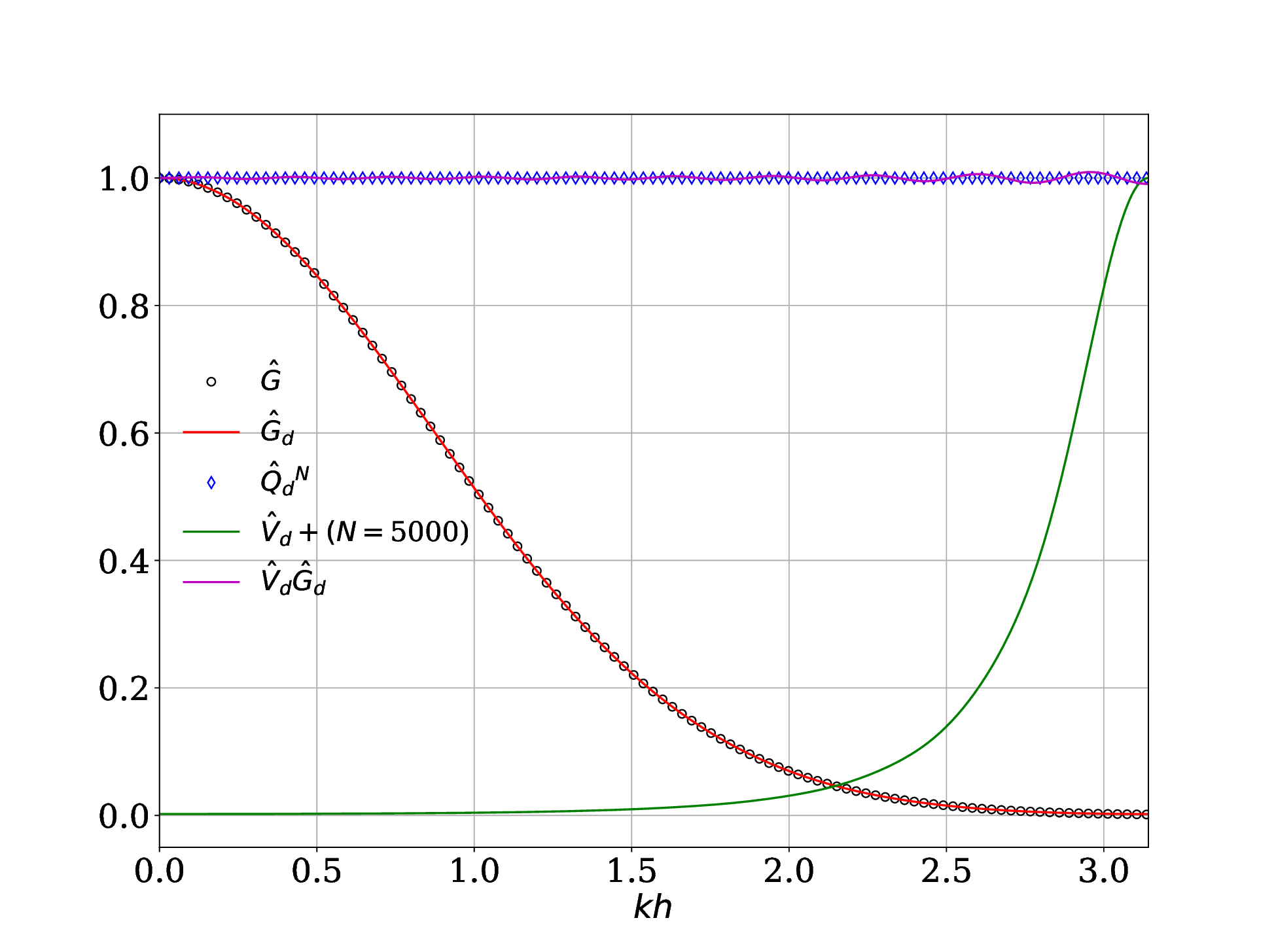}
}
\subfigure[]{
\includegraphics[width=0.55\textwidth, trim=3.0cm 0.0cm 0.0cm 0.0cm]{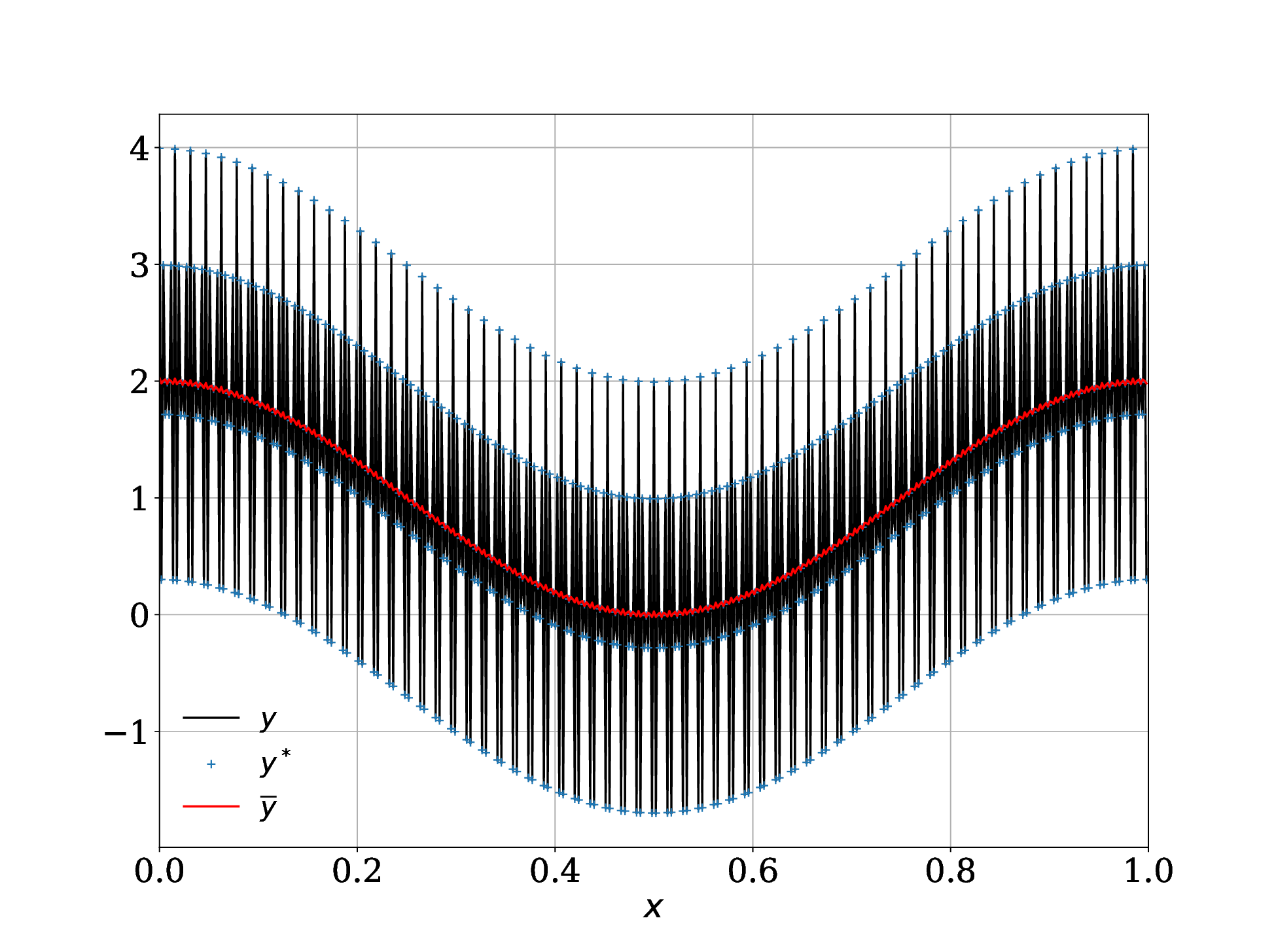}
}

\caption{Forward and inverse discrete filters for a Gaussian target transfer function and different reconstruction orders (a) $N=$50, (b) $N=$500, (c) $N=$5000. (d) shows the original signal $y$ (Eq. \ref{eq:lowPassTestSignal}), the filtered signal $\bar{y}$, and the reconstructed signal ${y^*}$  using the forward and inverse (deconvolution) filters designed using \LIBNAME\ whose transfer functions are shown  in (c).}
\label{fig:gaussianFI}
\end{figure}

As noted in the preceding sections, a major advantage and novelty of \LIBNAME\ is its ability to design inverse filters as well. The inverse filter design is based on van Cittert regularisation. The parameter $N$ in Eq. \ref{eq:regInverseTarget} determines the degree of reconstruction with larger $N$ corresponding to higher-order reconstruction (more wavenumbers recovered). In order to demonstrate this capability, forward and inverse filters are developed for a Gaussian target filter-a commonly used filter in CFD \cite{lele_jcp_1992, sagaut_ijnf_1999}. We choose the parameter of the filter $\Delta / h=4$ which results in $\hat{G}_d(k_ch)=0.5$ for $k_ch$=1.0197. The stencil size is calculated adaptively by setting the \textit{adaptive=True} keyword in the optimiser. The error for the forward filter design is set to $1.0 \cdot 10^{-6}$, and for the inverse filter to $1.0 \cdot 10^{-5}$. In addition, for the inverse filter we set the $monotonePos=True$ keyword to ensure that the inverse transfer function is monotonically increasing, and we set $\hat{V}_d(0)=1$. In order to test \LIBNAME\ for different reconstruction orders, we calculate inverse filters for $N=50$, $N=500$ and  $N=5000$.  The corresponding script for this test case is \textit{mainGaussianForwardAndInverse.py}. Figures \ref{fig:gaussianFI} (a)-(c) show the target transfer function of the Gaussian forward filter $\hat{G}$, the corresponding transfer function of the discrete filter $\hat{G}_d$ as obtained by running \LIBNAME, the target for the inverse filter design $\hat{Q}_d$, the inverse discrete filter transfer function $\hat{V}_d$, and finally the product $\hat{V}_d\hat{G}_d$ which is a measure of the performance of the inverse filter. Note that the inverse transfer function $\hat{V}_d$ in Figures \ref{fig:gaussianFI} (a)-(c) is normalised with respect to its maximum value. Table \ref{tbl:forwardAndInverseErrors} shows the corresponding error metrics, and the obtained stencil sizes for the forward and inverse filters i.e. $M_F$ and $M_I$ respectively.  It is clear from the results in Table \ref{tbl:forwardAndInverseErrors} that \LIBNAME\ performs well, with low mean-squared errors both for the forward and inverse filters. In addition, the maximum percentage overshoot in $\hat{V}_d\hat{G}_d$ remains small for all reconstruction orders. In a further test in physical space, the signal defined in Eq. \ref{eq:lowPassTestSignal} is filtered using the forward filter coefficients, and then reconstructed using the inverse filter coefficients as obtained using \LIBNAME. The transfer function of the forward and inverse filters is shown in Fig. \ref{fig:gaussianFI} (c) i.e. for a reconstruction order $N=5000$. Figure \ref{fig:gaussianFI} (d) shows the original signal, the filtered signal, and finally the reconstructed signal. It is clear that the inverse filter recovers the original signal down to the largest wavenumber component. Further 1D and 3D tests of the method, and using more complex signals having random wavenumber components can be found in our previous work \cite{nikolaou_caf_2023}.

\begin{table}[h!]
\centering
\footnotesize
\begin{tabular}{llllllll}
Target   & Parameter & $N$ & $M_F$
         & MSQE-F    & $M_I$ & MSQE-I     & Max \% overshoot $\hat{V}_d\hat{G}_d$ \\
\hline
Gaussian & $\Delta / h$ =4   & 50    & 4  & 1.2137E-08 & 9    & 7.8086E-06 & 1.1939E-01 \\
         &                   & 500   &  &              & 14   & 3.8127E-06 & 2.5627E-01   \\
         &                   & 5000  &  &              & 26   & 6.4389E-06 & 7.4747E-01 \\
\\
\hline
\end{tabular}
\caption{Errors for the forward and inverse discrete filters at different reconstruction orders $N$-the transfer function for the target Gaussian filter is given in Table \ref{tbl:filters}.}
\label{tbl:forwardAndInverseErrors}
\end{table}

\subsection{2D application: turbulent premixed flame, and images}

\begin{figure}[h!]
\centering
\includegraphics[width=0.55\textwidth, trim=0.0cm 0.0cm 0.0cm 0.0cm]{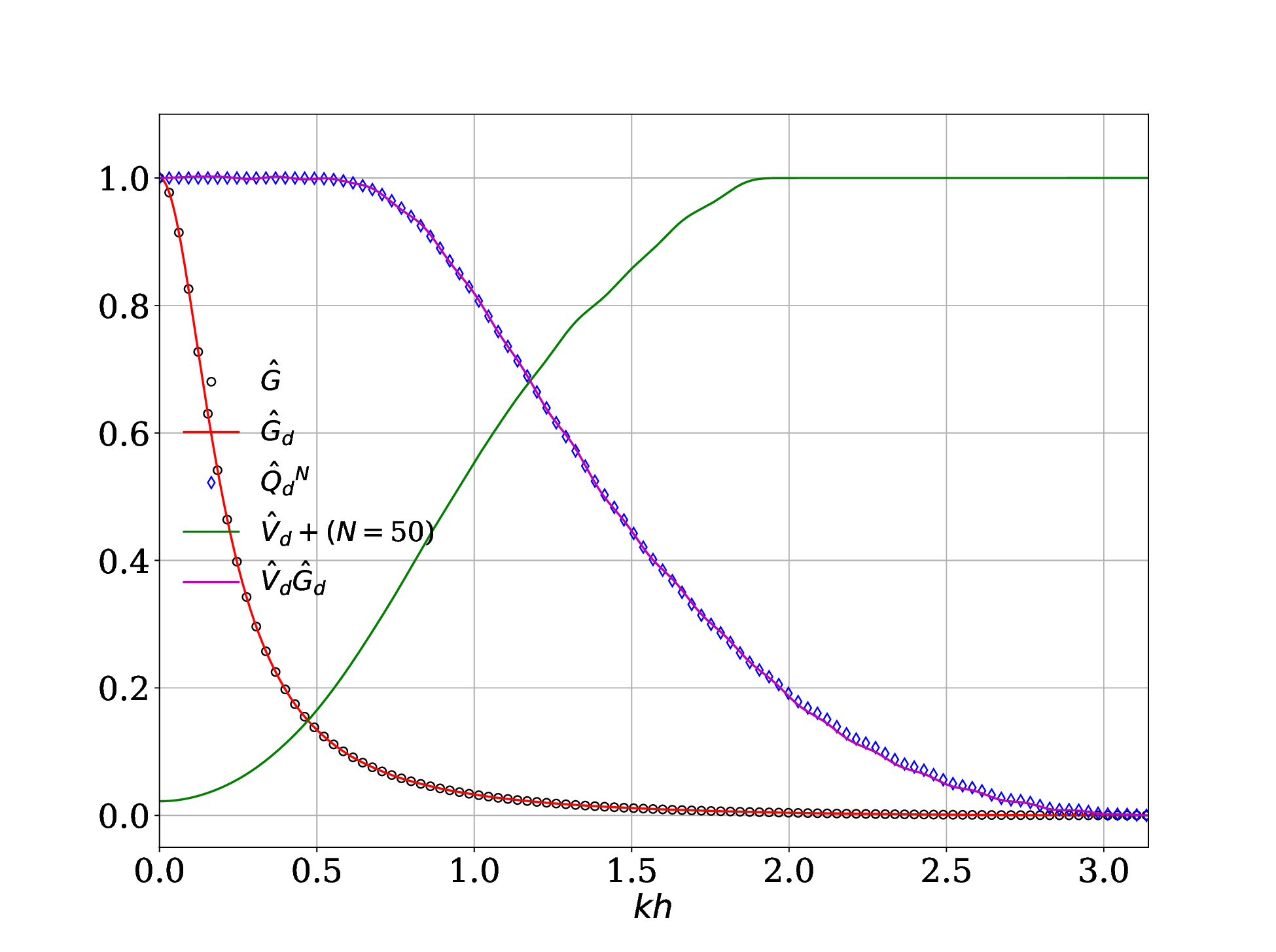}
\caption{Forward and inverse discrete filters for an Implicit target transfer function having parameter $a=-0.49$, and for $N=$50.}
\label{fig:3pointImplicit}
\end{figure}

\begin{figure}[h!]
\centering
\includegraphics[scale=0.40, trim=4.0cm 0.0cm 0.0cm 1.0cm]{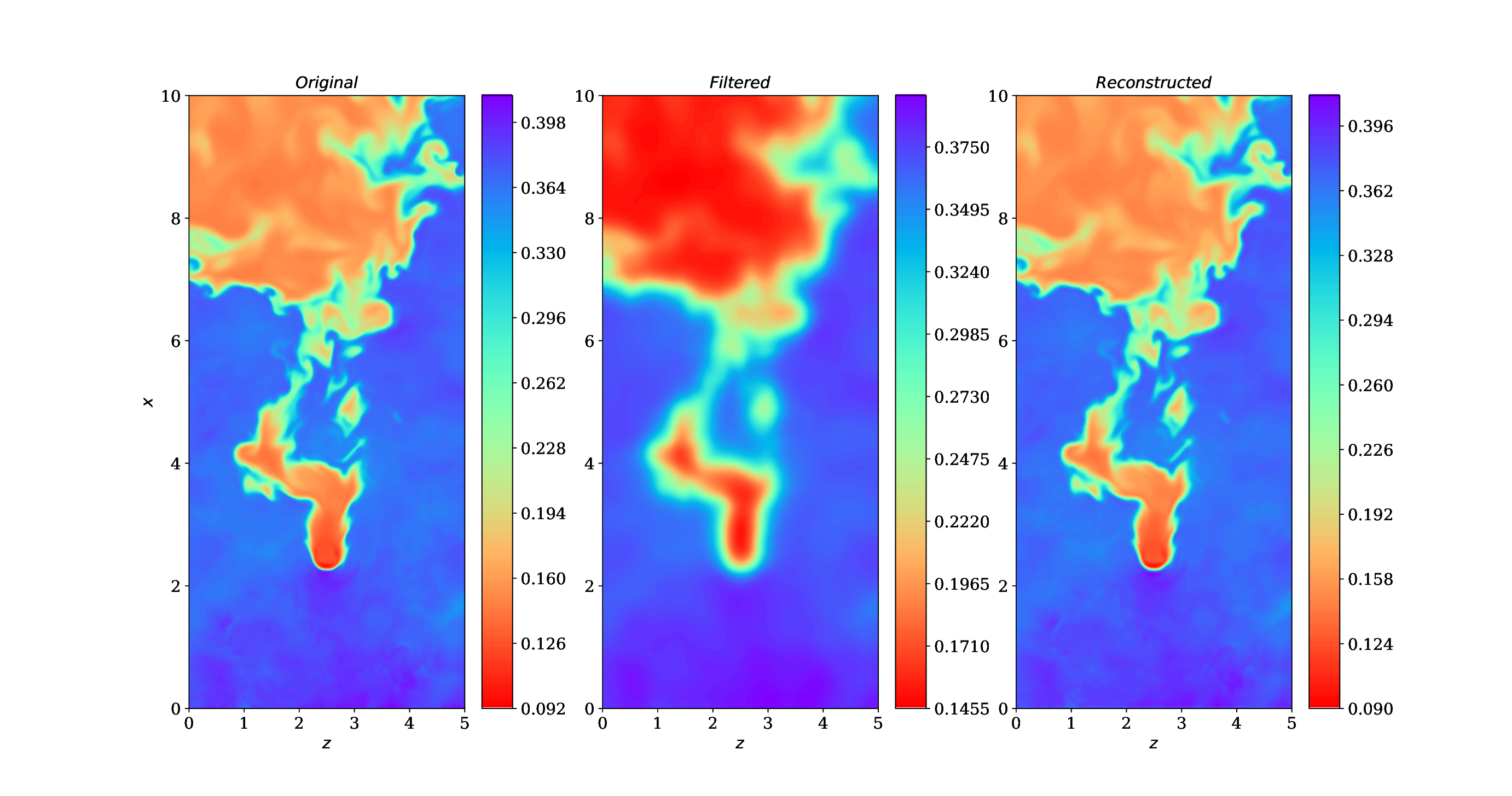}
\caption{The original, filtered, and reconstructed density field of a turbulent premixed V-flame anchored on a rod located at $x\simeq 5 mm$ downstream (dimensions are in mm): the turbulent inflow of reactants enters from the bottom of the domain, and the hot products exit from the top. The transfer functions of the forward and inverse filters empoyed are shown in Fig. \ref{fig:3pointImplicit}.}
\label{fig:turbflame}
\end{figure}

In this section, \LIBNAME\ is used to design forward and inverse 1D filters which are then extended to 2D using dimensional splitting \cite{sagaut_ijnf_1999}. The 2D forward and inverse (reconstruction) filters are then used to filter and reconstruct a highly turbulent signal. This signal corresponds to the density field of a turbulent premixed V-flame obtained using high-order direct numerical simulation. For further details of the simulation, numerical schemes used etc. we refer the reader to \cite{2024_jfm_nikolaou, 2011_minamoto_pof}. For this scenario, a 3-point implicit target transfer function is chosen having a parameter $a=-0.49$. This is an extreme value considering that the bounds for this parameter are $a \in (-0.5, 0.5)$. In order to keep the stencil size of the inverse filter within a reasonable range thus reducing the computational cost associated with 2D convolutions, we restrict the reconstruction order to $N=50$.
For the forward filter design, the optimisation error is set to $1.0 \cdot 10^{-8}$, and we set the constraints $\hat{G}_d(0)=1$, and $\hat{G}_d(\pi)=0$. For the inverse filter design, we set the optimisation error to $1.0 \cdot 10^{-5}$, $\hat{V}_d(0)=1$, and \textit{monotonePos=True}. 
 This choice is more than sufficient for practical applications in CFD \cite{nikolaou_caf_2023, 2024_jfm_nikolaou}. The corresponding script for this test case is \textit{main3PointImplicitForwardAndInverse.py}. The transfer function of the forward and inverse filters as obtained using \LIBNAME\ are shown in Fig. \ref{fig:3pointImplicit}. The corresponding stencil size of the forward filter is $M_F=79$ while for the inverse filter is $M_I=87$. Figure \ref{fig:turbflame} shows contours of the original density field, the filtered field as obtained by performing 2D convolution using the forward filter coefficients, and the reconstructed field as obtained by performing convolution on the filtered field using the inverse filter coefficients. Note that the filtering is substantial, and many of the fine-scale details of the turbulent flame are lost. Reconstruction recovers to a good extent many of the finer scale details as expected. 
 
\begin{figure}[h!]
\centering
\subfigure{
\includegraphics[scale=1.25, trim=1.5cm 1.0cm 0.0cm 0.0cm]{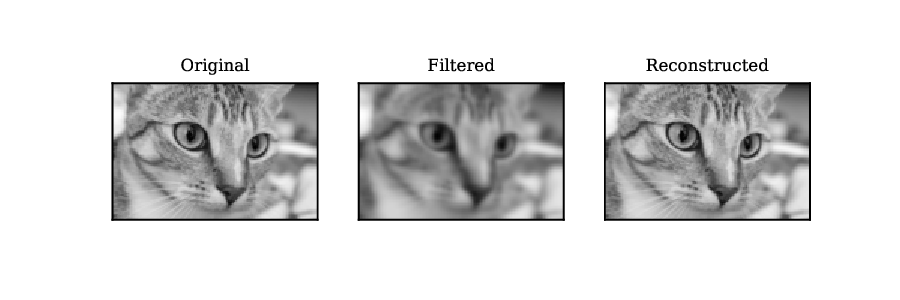}
}
\subfigure{
\includegraphics[scale=1.25, trim=1.5cm 0.0cm 0.0cm 0.0cm]{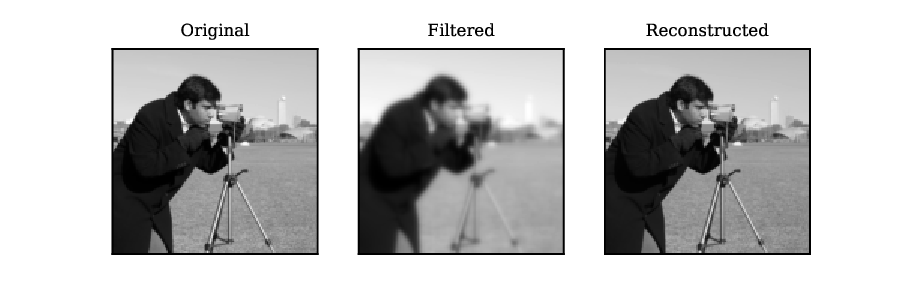}
}
\subfigure{
\includegraphics[scale=1.25, trim=1.5cm 0.0cm 0.0cm 0.0cm]{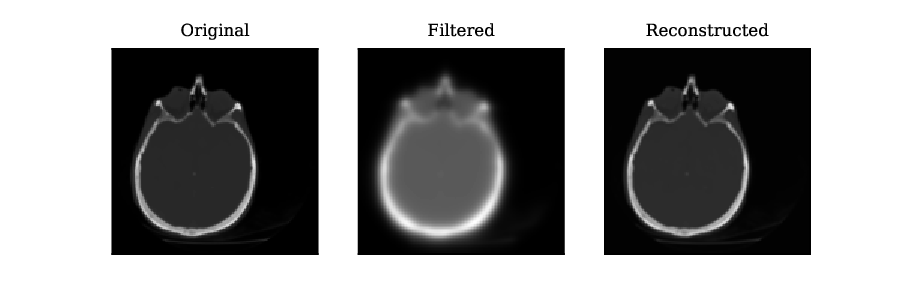}
}
\caption{Original, filtered, and reconstructed images: the transfer functions of the forward and inverse filters employed are shown in Fig. \ref{fig:3pointImplicit}.}
\label{fig:2dimages}
\end{figure}

In a further test, and to highlight the potentially large number of applications of \LIBNAME, the forward and inverse filters shown in Fig. \ref{fig:3pointImplicit} are used to filter and reconstruct a series of 2D images. The results are shown in Fig. \ref{fig:2dimages}. As with the turbulent flame case, filtering removes a substantial amount of information from the images, while reconstruction even at the modest degree of $N=50$, recovers almost  completely the original image. Further tests of the method including full 3D fields on large-scale computational meshes can be found in our earlier work \cite{nikolaou_caf_2023, 2024_jfm_nikolaou}. 

\clearpage
\section{Conclusions}

In this work, an open-source Python software library was developed for designing forward and inverse discrete filters. \LIBNAME\ is a general, abstract, and domain-specific numerical framework based on constrained optimisation which allows the user to design symmetric discrete filters subject to a wide range of user-defined transfer function constraints. In addition, \LIBNAME\ includes an adaptive filter stencil size computation, and a user-defined inverse filter design reconstruction order. These capabilities, make \LIBNAME\ unique in its class for designing with ease a wide range of discrete filters including low/high-pass, multi band-pass, multi band-stop to name but a few. Due to its generality and abstraction, the user may well define their own target transfer function for which to develop forward and inverse filters. \LIBNAME\ has been extensively validated in this work against benchmark classic methods in the literature (Kaiser window, Parks-McClellan) for forward filter design as well as for designing a wide range of inverse filters. Although developed primarily for designing discrete filters for application to computational fluid dynamics simulations, and deconvolution-based modelling, \LIBNAME\ can be used in a plethora of signal-processing and image-processing applications.

\section*{Authors contribution}

Zacharias Nikolaou: Conceptualization, Methodology, Software, Writing-review \& editing, Writing-original draft, Visualization, Validation, Data curation. P. Domingo: Writing-review \& editing, Writing-original draft. Luc Vervisch: Writing-review \& editing, Writing-original draft. D. Drikakis: Writing-review \& editing, Writing-original draft.

\section*{Declaration of competing interest}

The authors declare that they have no known competing financial
interests or personal relationships that could have appeared to influence the work reported in this paper.

\bibliographystyle{unsrt}
\bibliography{ref.bib}

\end{document}